# Phase Transitions on the Surface of a Carbon Nanotube

Zenghui Wang, Jiang Wei, Peter Morse, J. Gregory Dash, Oscar E. Vilches and David H. Cobden*
Department of Physics, University of Washington, Seattle WA 98195-1560

*cobden@u.washington.edu

**A suspended carbon nanotube can act as a nanoscale resonator with remarkable electromechanical properties[1-4] and the ability to detect adsorption on its surface at the level of single atoms[5-7]. Understanding adsorption on nanotubes and other graphitic materials is key to many sensing and storage applications. Here we show that nanotube resonators offer a powerful new means of investigating fundamental aspects of adsorption on carbon, including the collective behaviour of adsorbed matter and its coupling to the substrate electrons. By monitoring the vibrational resonance frequency in the presence of noble gases, we observe the formation of monolayers on the cylindrical surface and phase transitions within these monolayers, and simultaneous modification of the electrical conductance. The monolayer observations also demonstrate the possibility of studying the fundamental behaviour of matter in cylindrical geometry.**

Carbon nanotubes have great potential for gas sensing[8], since they present a seamless graphene surface to the environment and their excellent electrical properties are governed by electrons moving on the surface. However, the mechanisms behind the often surprisingly large sensitivity observed in nanotube ensembles[9,10] and in conventional nanotube devices on insulating substrates[11] remain obscure. More generally, a better understanding of adsorption of various substances on graphitic materials is important for many technologies including lithium batteries, and gas purification and storage. Adsorbated films exhibit many kinds of ordering, and decades of studies[12] on bulk exfoliated graphite have revealed a huge range of two-dimensional (2D) phenomena within the first adsorbed layer, including 2D melting, transitions between solids which are either commensurate or incommensurate with the graphene lattice, and critical behaviour.

Here we demonstrate that suspended nanotube resonators provide a revolutionary approach to the study of adsorption on graphitic materials. Their combination of remarkable electrical and electromechanical properties[2-4], clean geometry, and single-atom mass sensitivity[5-7], means that one can simultaneously measure both the precise amount of adsorbed substance and its effect on the electrical properties for a single, isolated, pristine nanotube. As a result one can not only explore the collective behaviour of the adsorbates but also quantify the interaction between adsorbates and substrate electrons. For simplicity we focus on noble gases, in which we observe monolayer formation and dramatic phase transitions, together with signatures of phase transitions in the conductance. Our results also demonstrate intriguing possibilities for exploring ordering with periodic boundary conditions and approaching the one-dimensional (1D) limit, avoiding problems of heterogeneity that have complicated previous explorations of this regime using conventional techniques which require the use of bulk nanotube samples[13-15].

Fig. 1a is an electron micrograph of a suspended nanotube device, made by pre-fabricating the electrodes and trench and growing the nanotubes in the last step[16] to avoid any chemical exposure which might contaminate the pristine nanotube surface. Mechanical resonances were detected using the mixing technique[1], where they appear as sharp features in mixing current signal during frequency sweeps. The frequency $f_{res}$ of each resonance decreases with increasing pressure $P$ as gas adsorbs, as illustrated in Fig. 1b for Kr at 77 K. Equilibrium is established in seconds after a small pressure change (see Fig. 1c), and $f_{res}$ is a reproducible function of $P$ (see Fig. 1d). We deduce the adsorbed mass by assuming (this will be justified below) that at a given temperature $f_{res}$ varies as $\rho^{-1/2}$, where the total mass per unit length $\rho$ is the sum of that of the bare nanotube, $\rho_0$, and of the adsorbates, $\Delta\rho$. The fractional mass increase of the nanobalance is then $\Delta\rho/\rho_0 = (f_0/f_{res})^2 - 1$, where $f_0 = \lim_{P \to 0} f_{res}$. Since $f_0$ is determined separately at each temperature $T$, any variation of the resonance frequency with $T$, such as might result from thermal expansion, is factored out. If $m_C$ and $m_{ads}$ are the atomic masses of carbon and the adsorbed species respectively, then the quantity

$$\phi = (\Delta\rho/m_{ads})/(\rho_0/m_C) = m_C/m_{ads} \left[(f_0/f_{res})^2 - 1\right], \qquad (1)$$

is the number of adsorbed atoms per carbon atom. Fig. 1e shows examples of isotherms of $\phi$ vs $P$ derived in this way. (Further experimental details are given in the supplementary information.)

The assumed $\rho^{-1/2}$ scaling of $f_{res}$ requires that the change in elastic properties be negligible compared with the fractional change in mass. This is justified because the covalent C-C bond is two orders of magnitude stronger than the van der Waals attraction between adsorbates. It also requires that the adsorbed mass be distributed uniformly over the nanotube surface. This would not be the case if part of the surface were contaminated, or if a denser phase appeared preferentially at the ends or in the middle due to long-range forces. Nonuniformity would cause different vibrational modes to shift in different ways. However, we have found that $f_0/f_{res}$ does not depend on which mode is used, and also that it is insensitive to $V_g$ (see Fig. 1e). Taken together with the results described below, these observations indicate that many of our devices consist of single-walled nanotubes in which $\phi$ is a good measure of the coverage (number of adsorbates per surface atom), and that the coverage is uniform.

Isotherms of $\phi$ vs $P$ for argon are dominated by a large, smooth step, as shown in Fig. 2 for device YB3. A similar step is well known in conventional volumetric isotherms on bulk exfoliated graphite, and is characteristic of the densification of a supercritical 2D fluid (F), which occurs above 56 K, the 2D critical point of Ar on graphite[12,17,18]. The colder Ar isotherms also show a second, smaller step at $\phi \approx 0.24$ which is focused on in the inset. A similar step is seen graphite[18] when the fluid freezes to a 2D incommensurate solid (IS). Accordingly we anticipate an F+IS coexistence region at lower temperatures of the form indicated by the dashed lines. We note however that corresponding features occur at higher pressures on nanotubes than on graphite. This is in keeping with the expected weaker binding of atoms to a nanotube surface than to bulk graphite.

Krypton, with its greater size and polarizability, provides an instructive contrast with argon. Isotherms of Kr on device YB3 exhibit a dramatic vertical step followed by two smaller steps (Fig. 3.). Again, these resemble conventional volumetric isotherms of the same substance on exfoliated graphite[19] but shifted to higher pressures. The size and sharpness of the large step implies not only a first-order phase transition but also excellent substrate homogeneity, consistent with there being no grain boundaries or imperfections on surface of a single nanotube. The first plateau, between the first two steps, is narrow and easily missed: it is resolved in the 77.4 K data here but not the 73.7 K data.

Whereas Ar does not form any commensurate phase on graphite, Kr condenses from a low-density 2D vapor (V) to a commensurate solid (CS) with one Kr atom per six C atoms[20], as indicated in the left inset to Fig. 3. At higher pressure it converts to an IS[21]. Remarkably, the first plateau in our Kr isotherms whenever resolved is centered at $\phi = 1/6$, corresponding precisely to the coverage of the commensurate solid and implying that the first step is a V-CS transition. The highest plateau reached is likely to be the IS, while the intermediate plateau is of unknown nature but may be related to the proposed reentrant fluid phase[22].

The fact that $\phi$ takes the value of 1/6 for the commensurate phase strongly indicates that the assumption behind Eq. (1) that $f_{\text{res}} \propto \rho^{-1/2}$ is justified for this device. Some other devices, including YB1 (see Fig. 1g), showed similar behaviour but relatively small values of $\phi$. This can be explained if these devices have a reduced ratio of available surface area to nanotube mass, which would be the case for a multi-walled nanotube or if part of the nanotube's surface were contaminated.

The identification of commensurate and incommensurate 2D solids on the cylindrical nanotube surface raises many interesting questions. The solid is subject both to the curvature, which breaks the isotropy of the graphene lattice, and to the cylindrical boundary condition. It may be rolled seamlessly like the underlying graphene, or alternately it may contain a domain wall running along the nanotube. For the CS, which is in registry with the carbon surface, the seamless case occurs only when $(N-M)/3$ is an integer, where $(N,M)$ is the nanotube's roll-up vector. Interestingly, this is precisely the same condition as for the nanotube to be metallic[23]. From the $V_g$ dependence of its conductance $G$ (right inset to Fig. 3) we identify YB3 as a small-gap metallic nanotube which obeys this condition.

Not only do these nanotubes show beautiful mass adsorption characteristics but also their electrical properties can be measured at the same time, allowing exploration of the coupling between adsorbates and electrons which is impossible in conventional adsorption experiments and which is central to sensing applications. Fig. 4 shows the $\phi$–$P$ isotherm of Kr at 77 K for another device, YB8. It exhibits a large sharp step at about 16 mTorr similar to that in to YB3. The upper left inset shows $G$-$V_g$ characteristics measured at pressures on either side of the phase transition. The conductance is clearly suppressed at positive $V_g$ at the higher pressure. The other trace in the main panel shows the resistance measured at a fixed positive gate voltage. It increases gradually at low coverages and jumps suddenly at the transition. This behaviour is reproducible and reversible.

One immediate consequence of this observed coupling is that we can investigate for the first time the dynamics of such a phase transition. The right inset to Fig. 4 shows the resistance monitored with a 10 ms instrumental response time as the Kr pressure is increased rapidly across the transition. The step occurs in about 0.1 second, indicating that this is the intrinsic time scale of the phase transition on the nanotube.

The $G$-$V_g$ characteristics imply that this nanotube has a small gap (~60 meV) and the conductance is limited by tunneling across the gap at positive $V_g$ but not at negative $V_g$ (see supplementary information). The larger decrease in conductance at the phase transition for positive $V_g$ could therefore be explained by an increase in the gap. It is interesting to note that the CS has a reciprocal lattice vector which connects the Dirac points in the graphene Brillouin zone, and hence coherent scattering from commensurate Kr offers a possible mechanism for modifying the gap. We note that the Kr atoms will be statically polarized by the gate-induced electric field perpendicular to the nanotube surface, but because of the relatively weak field in our device geometry we do not expect this polarization to have significant consequences (see supplementary information). Further experiments of this type should yield many more important insights into the coupling not only of noble gases but of many other substances to electrons in graphitic carbon.

**Acknowledgments**
We acknowledge useful discussions with Marc Bockrath, Hsin-Ying Chiu, Milton W. Cole,

Marcel den Nijs, Michael Schick, and Arend van der Zande. This work was partly supported by the NSF, the PRF, the UW Royalty Research Fund, and a UW UIF fellowship. Portions of this work were done in the UCSB Nanofabrication Facility and in the UW Nanotechnology Center, which are parts of the NSF-funded NNIN network.

**Figures**

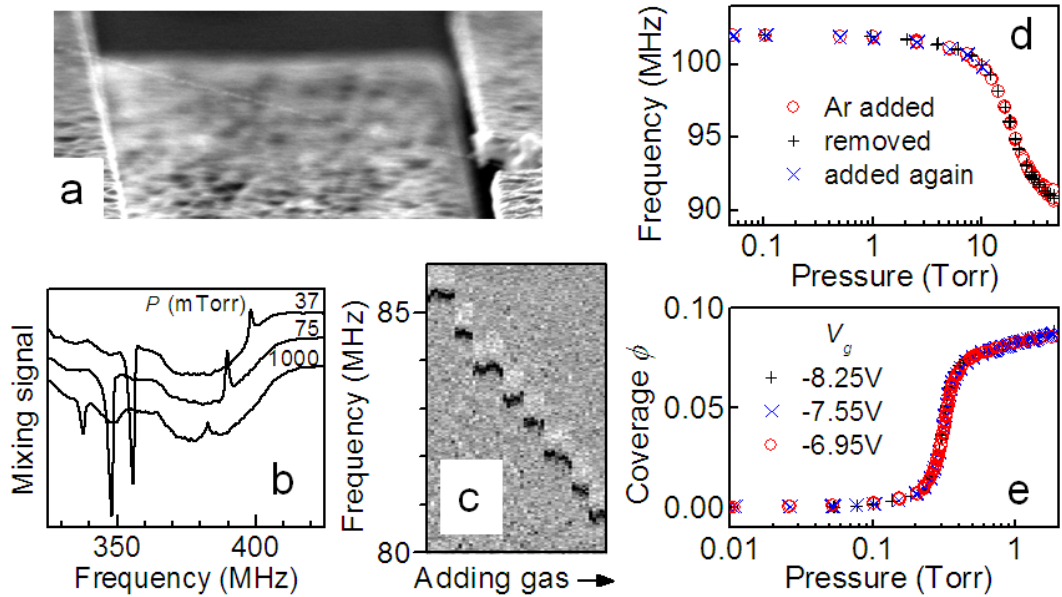

**Fig. 1.** Detecting adsorption on a single vibrating nanotube. (a) Electron micrograph of a nanotube spanning a 2-μm wide trench between Pt contacts. (b) Line traces of the mixing current signal showing resonances shifting with Kr gas pressure for device YB3 (0.5 μm gap) at 77.4 K and $V_g$ = 9.58 V. (c) Grayscale of the mixing signal for a series of frequency sweeps taken as the Kr pressure is increased stepwise every few minutes, demonstrating rapid equilibration, for device YB1 (2 μm gap) at 77.4 K. (d) Variation of resonance frequency with Ar pressure for YB1, showing high reproducibility. (e) Variation of coverage parameter $\phi$ obtained using Eq. (1) with Kr pressure, for YB1 demonstrating little sensitivity to gate voltage.

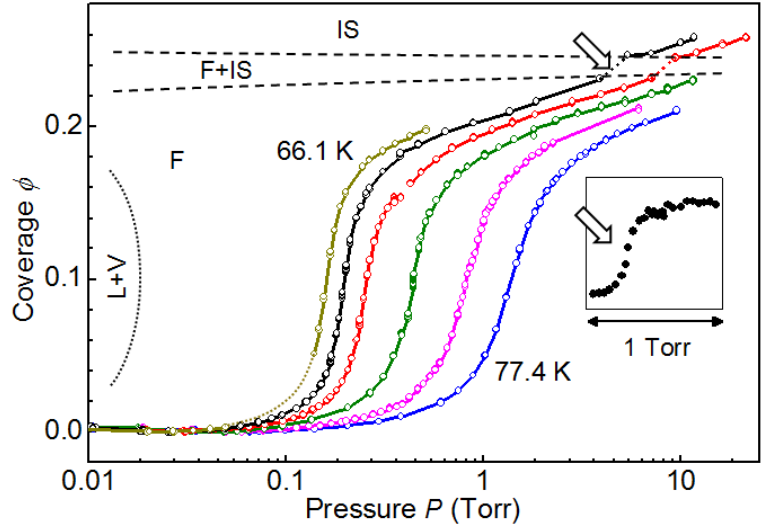

**Fig. 2.** Isotherms of coverage parameter $\phi$ for Ar on device YB3. The temperatures are 66.1, 67.7, 68.8, 71.0, 73.9 and 77.4 K. The large step occurs within the supercritical fluid F, with a first-order liquid-vapor (L-V) transition expected below a critical point at ~ 0.01 Torr. The smaller step, a separate measurement of which is shown in the inset, occurs on transition to an incommensurate solid (IS). Dotted and dashed lines indicate boundaries of coexistence regions.

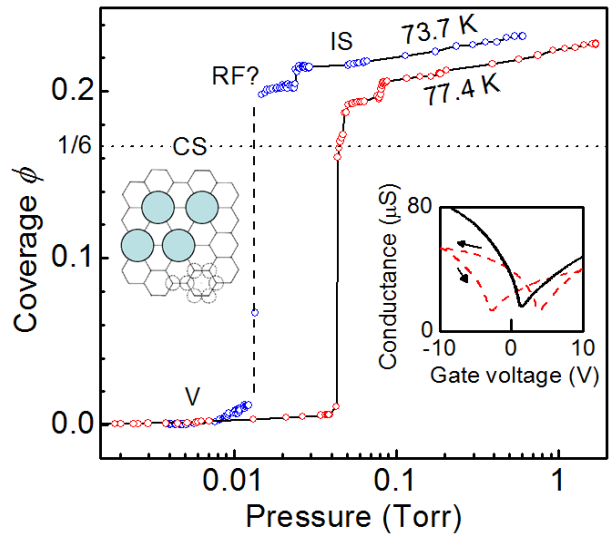

**Fig. 3.** Adsorption of Kr on device YB3, showing first order phase transitions. (The 73.7 K isotherm has incomplete data at the dashed line.) The dotted horizontal line at $\phi = 1/6$ corresponds to the expected coverage of 1 adsorbate per 6 carbon atoms in a commensurate solid on a clean single-walled nanotube. The left inset shows the commensurate arrangement of adsorbates (shaded circles) sitting on every third carbon hexagon. The right inset shows conductance vs. gate voltage (solid: vacuum; dashed: in air) at room temperature, whose form is that of a small-bandgap nanotube.

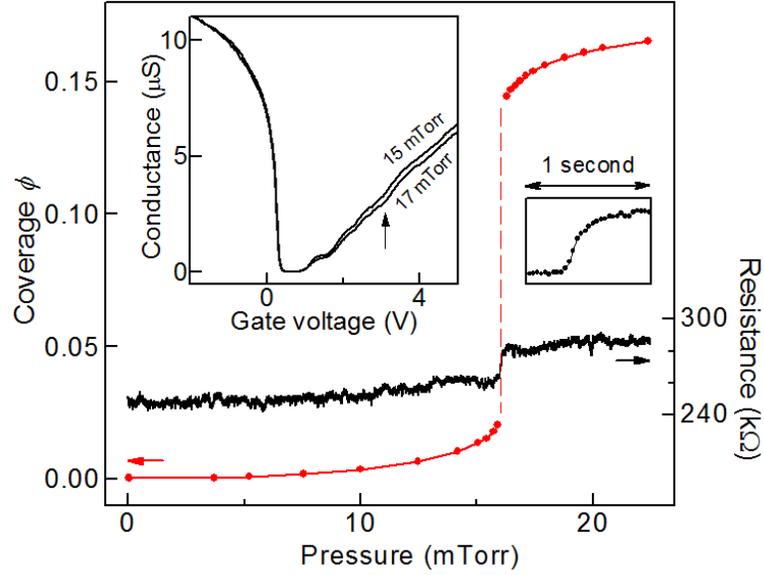

**Fig. 4.** Combined mass and transport measurements on device YB8 (1 µm gap) exposed to Kr at 77 K. Main: measurements of both coverage parameter $\phi$ (left axis) and resistance $R$ (right axis) as a function of pressure. Left inset: conductance vs. gate voltage at pressures just below (15 mTorr) and above (17 mTorr) the transition, with an arrow indicating the gate voltage of +3.1 V at which $R$ was measured. Right inset: resistance vs. time during a rapid upward pressure sweep across the transition.

# Phase Transitions on the Surface of a Carbon Nanotube: Supplementary information

Zenghui Wang, Jiang Wei, Peter Morse, J. Gregory Dash, Oscar E. Vilches, and David H. Cobden

Department of Physics, University of Washington, Seattle WA 98195-1560

**Device fabrication**

The suspended nanotube devices were made using a technique based on that of Cao et al.[1] Arrays of Pt/Cr contacts were formed on the sides of trenches 0.5 – 2 μm wide and 300 nm deep, etched in a silicon wafer coated with 500 nm of $SiO_2$ topped by 60 nm of $Si_3N_4$. The trench bottoms were also coated with Pt. The nanotubes were grown by chemical vapor deposition from catalyst pads adjacent to the trench using the same parameters as Kong et al[2]. Most should be single-walled with diameters in the range 1-3 nm. After growth no further processing was done. Scanning electron microscopy was not used on the best working devices to avoid damage. The devices were kept in vacuum to minimize surface contamination, then were wire-bonded into ceramic packages and mounted in a cell to allow evacuation and exposure to low pressure gases.

**Resonance detection**

Vibrational resonances were detected using a simplified version of the mixing technique invented by Sazonova et al.[3] As indicated in Fig. S1, an ac voltage $V_s$ at frequency $f$ (1-500 MHz) with 99% amplitude modulation at $f_{mod}$ = 1 kHz is applied to the source electrode, with the drain connected to a virtual-earth current preamplifier. A dc gate voltage $V_g$ is applied to the heavily doped Si substrate. The ac voltage difference between the nanotube and the gate underneath generates an oscillation in the electrostatic attraction on the nanotube, larger near the source, which drives the nanotube to vibrate at the same frequency. As the conductance of the nanotube at a given $V_g$ depends on the capacitance between the nanotube and the gate electrode, mechanical motion of the nanotube at frequency $f$ relative to the gate beneath causes a corresponding oscillation of its conductance. Combined with the source bias components at $f \pm f_{mod}$ this generates a mixing signal at $f_{mod}$ in the current, which is detected using a lock-in amplifier referenced to the $f_{mod}$ signal. A vibrational resonance appears as a sharp feature in the lock-in output when sweeping $f$. Fig. S2 shows a device (YB3) in vacuum with two visible resonances. Their positions increase roughly quadratically with $V_g$, consistent with the effect of electrostatic attraction towards the gate. Adsorption measurements were performed at values of $V_g$ where a resonance stood out most clearly from the $V_g$-independent background. We typically used rms $V_s$ values between 5 and 15 mV rms, after determining that the magnitude of $V_s$ did not affect the resonance position at 77 K. The low-frequency conductance was measured by applying an additional ac source-drain excitation of 1 mV rms at 250 Hz and detecting the resulting current component in the output of the same preamplifier with a second lock-in amplifier.

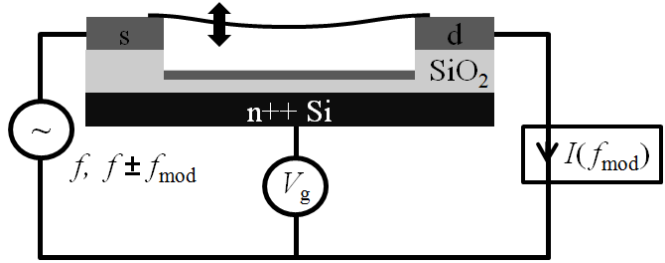

Fig. S1. Schematic measurement setup. The source s, drain d, and bottom of etched trench are coated in Pt (dark gray). The mixing signal at the modulation frequency, $I(f_{mod})$, depends on the amplitude of oscillation of the nanotube at carrier frequency $f$. A low frequency ac source bias (not shown) is applied at the same time to measure the conductance of the nanotube.

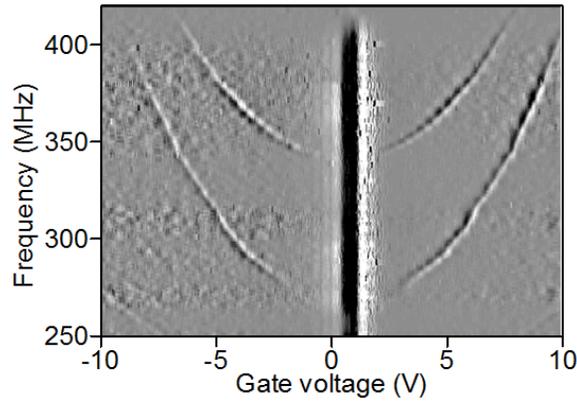

Fig. S2. Grayscale plot of mixing signal vs. driving frequency and gate voltage in vacuum at 300 K for device YB3. A smoothed background was subtracted. Two resonant modes can be clearly seen.

**Isotherm measurement**

The cell holding the device is connected to a gas handling system with capacitance pressure gauges. The cell is submerged in liquid nitrogen at 77.4 K, which can be further cooled by up to 11 K by bubbling helium gas through it. The usual procedure for obtaining an isotherm of $\phi$ vs. $P$ is as follows. With the temperature stabilized, the device is first "reset" by turbo pumping to remove any previously adsorbed gas. Then gas is introduced to the cell either in discrete shots (producing step-like increases in pressure) or through continuous slow bleeding (producing a smooth increase in pressure) in such a way that the system is always near equilibrium. The generator frequency $f$ is repeatedly swept and in each sweep the position of the resonance is identified together with the corresponding pressure on the gauge. The pressure $P$ in the cell is deduced from the gauge pressure taking into account a thermomolecular correction[4]. After each pressure sweep is complete, $f_0 = \lim_{P \to 0} f_{res}$ is determined and the frequency values are converted to $\phi$ values using Eq. 1 in the main text.

## Interpretation of $G$-$V_g$ characteristics

In the $G$-$V_g$ characteristic of device YB8 (Fig. 4, left inset) the first few Coulomb oscillations can be seen. We estimate the gate coupling efficiency $X$ by assuming that the 400 mV width of the Coulomb oscillations in $V_g$ is determined by thermal broadening (corresponding 3.5 $k_BT \approx$ 23 meV at 77 K) and that each corresponds to a single added electron, yielding $X \approx 23/400 \approx$ 0.06. The size of the band gap corresponds roughly to the ~1 V gate voltage difference between the steepest point of the p-type transconductance and the first n-type Coulomb peak, giving $E_g \sim$ (1 V)$eX \sim$ 60 meV.

The fact that the p-type conductance is higher, that Coulomb oscillations are only seen on the n-type side, and that the zero conductance region is centered at $V_g \sim$ +0.6 V, can be explained by the combination of a small gap, the higher work function of platinum than carbon, and high contact transparency. At negative $V_g$ there is then barrier-free conduction through the valence band of the nanotube, while at positive $V_g$ electrons must tunnel across the nanotube's band gap in and out of a Coulomb island in the conduction band, as sketched in Fig. S3.

From the period of the Coulomb oscillations we see that the charge $Q$ on the island is approximately one electron per positive volt on the gate. Very similar devices in the literature show similar parameters[5]. As a result we may estimate the perpendicular electric field at the surface of the nanotube roughly as $E \sim Q/(\pi \varepsilon_0 L d)$ where $d$ is the nanotube diameter and $L$ the length, giving $E \sim 5 \times 10^7$ V/m for $V_g$ = 10 V. The atomic polarizability of Kr is $\alpha$ = 2.48, so its dipole energy is $\sim 4\, \pi\varepsilon_0\alpha E^2 \cdot (1\, \text{Å})^3 \sim$ 4 μeV at $V_g$ = 10 V, which is three orders of magnitude less than $k_BT$ and hence probably insignificant.

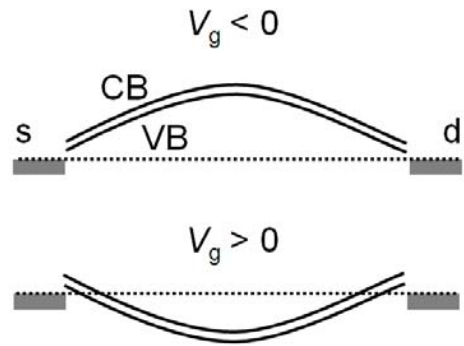

Fig. S3. Sketched band diagrams of a small-gap nanotube at negative and positive gate voltage, indicating how conduction in the latter case requires Zener tunneling between valance band (VB) and conduction band (CB) and is thus sensitive to the band gap.